\documentclass[epsf,12pt]{article}
\usepackage{amsmath}
\usepackage{amssymb}
\usepackage{graphicx}
\usepackage{geometry}
\date{}

\title{
The number of gauge singlets in supersymmetric Yang-Mills quantum mechanics}
\author{Maciej Trzetrzelewski
\footnote{trzetrzelewski@th.if.uj.edu.pl }   \\
 \emph{M. Smoluchowski Institute of Physics, Jagiellonian
University} \\
\emph{Reymonta 4, 30-059 Krak\'ow, Poland}
}

\begin{document}
\maketitle

\abstract{ We compute generating functions for number  of
$U(N)$($SU(N)$) singlets in Fock space in several space dimensions.
The motivation to find the explicit form of the functions is from
the numerical approach to supersymmetric Yang-Mills quantum
mechanics, based on Fock space. Incidently the functions give many
important insights into the quantum mechanical models based
$U(N)$($SU(N)$) gauge group. }

\section{Introduction}

Yang Mills theories in zero volume limit provide a natural arena
where quantum mechanics with the singlet constraint emerge due to
the Gauss law \cite{Luscher}. The gauge fields $A^a_{i}$, where $i$
and $a$ are spatial and color indices respectively, become now the
coordinate operators $x^a_i$ in emerging Yang-Mills quantum
mechanics (YMQM). The resulting system is extremely difficult to
solve and accordingly there are no exact solutions in the
literature.

The model and its supersymmetric extension (SYMQM) play an important
role in quantum mechanical description of the membrane \cite{Hoppe}
and the supermembrane \cite{Bergshoeff}, i.e. they give a
regularized description of a (super)membrane. The $D=9+1$
dimensional SYMQM model \cite{Claudson} became famous due
to the BFSS conjecture \cite{Banks}, relating the later to M-theory,
and since then has been studied in number of papers ( we refer to the
existing reviews  \cite{rev} ). However there are no nonperturbative
calculations of e.g. the spectrum of BFSS matrix model because of
the high complexity ( of e.g. the Fock space ) that makes any
numerical approach difficult.

A particularly useful approach is the cutoff method \cite{WosiekD4},
which consists of representing the hamiltonian in truncated Fock
space and diagonalizing the resulting finite matrix. In the number
of papers \cite{WosiekD4,WosiekD2,WosiekD41,vanBall,Kotanski} the
method, applied to the $D=2,4$, $SU(2)$ system, proved to be very
fruitful giving the nonperturbative values of the spectra, wave
functions, the Witten index etc. The Fock space approach can also be
applied to the systems with large number of colors $N$
\cite{VW,VW1,Onofri,WVspin,WVspin1} where the nonperturbative
results may shed some new light into the area of large $N$ quantum
field theory.

An important step in applying the cutoff method is the construction
of the $SU(N)$ invariant basis in Fock space. This however proved to
be very time consuming in some cases \cite{WosiekD10}. We focus in this
paper on computing the number of $SU(N)$ singlets in Fock space
analytically hence facilitating the numerical considerations. The
difficulty in computing this number arises due to many identities (
in fact an infinite number of them, emerging from the Cayley-Hamilton
theorem ) among $SU(N)$ singlets that make them linearly  dependent.
There are, of course, very well known algorithms to produce the
singlet states \cite{Dittner}. One simply takes the trace operators,
e.g. $Tr(x_ix_j\ldots)$, $x_i=x^a_iT_a$ where $T_a$'s are $SU(N)$
generators in the fundamental representation, and act with an arbitrary
number of products of trace operators on e.g. the Fock vacuum. Such
set of states certainly spans the whole Fock space however they are
in general linearly dependent. One can use the Cayley-Hamilton
theorem to find the linearly independent ones but this is in general
tedious when the number of $x_i$ matrices becomes large.

 There is
however a way to circumvent this problem when $N=\infty$ since, in that case, there are no additional identities among singlets. Furthermore, if one is interested
only in single trace states then the famous Polya theorem \cite{Polya} can be applied to compute the
number of singlets explicitly. The first computation of this kind is
by Sundborg \cite{Sundborg}, Polyakov \cite{Polyakov}  and Aharony et al. \cite{Aharony} in the
context of weakly coupled Yang Mills theories and by Semenoff et al.
\cite{Semenoff} in the context of BFSS matrix model.

In quantum mechanical systems the property, that
the single trace states are the most important ones in large $N$
limit, is not always valid. One can verify \cite{VW} that in the case of the anharmonic oscillator
the single trace states do not reproduce the correct \cite{Brezin}
large $N$ spectrum. A different example is the free hamiltonian
\cite{33largeN} where it was explicitly shown that all the $1/N^k$
terms play an important role in the large $N$ limit due to the
distinguished role of the bilinear operators $Tr(x^2)$.

In this paper we neither assume that $N$ is large nor do we consider
only the single trace states. The method that we will exploit in
details in sections 2 and 3 is the very well known character method
in group theory\footnote{I thank R. Janik for the idea.}. In this framework we compute the generating
functions for numbers of singlets in sectors with given number of
bosonic/fermionic quanta and therefore give the exact number of
singlets. In section 2 we discuss only the $D=2$ SYMQM. The
generating function of number of gauge singlets can be related, in
this case,  to the Witten index of particular models. In this way we
obtain some physical constraints, on seemingly unrelated group
theory numbers, which we find very interesting. In section 3 and 4
we do the analogous computation only for $D=4,6,10$ and relate our
generating functions with the partition function of the weakly
coupled Yang-Mills theories on $R\times S^1$.

While this work was done a related paper \cite{Dolan} by F. Dolan
was published. There is a significant overlap between \cite{Dolan}
and our paper although the motivation as well as the detailed
calculations are in different spirit.

\section{The character method for $D=1+1$ SYMQM}

Let us consider the most general state in
$D=1+1$ SYMQM

\[
\mid s \rangle = T_{b_1 \ldots b_{n_B}c_1 \ldots
c_{n_F}}{a^{\dagger}}^{b_1}\ldots
{a^{\dagger}}^{b_{n_B}}{f^{\dagger}}^{c_1}\ldots
{f^{\dagger}}^{c_{n_F}}\mid 0 \rangle \label{T}.
\]
Here $\mid 0 \rangle$ is the Fock vacuum while $a^{\dagger \ b}$ and
$f^{\dagger \ b}$ are bosonic and fermionic creation operators being
in the adjoint representation of the Lie group G. The operators obey
the commutation  and anticommutation relations $[a^b,a^{\dagger \
c}]=\delta^{bc}$, $\{f^b,f^{\dagger \ c}\}=\delta^{bc}$. The tensor
$ T_{b_1 \ldots b_{n_B}c_1 \ldots
c_{n_F}}$ is a group invariant tensor so that the state $\mid s
\rangle$ is a group singlet.
 The number of bosonic quanta in $\mid s \rangle $ is
$n_B$  and the number of fermionic quanta is  $n_F$. In this case we
say that $\mid s \rangle $ is in the $(n_B,n_F)$ sector.

 For the Lie group G we take
$U(N)$ ( or $SU(N)$ ). The adjoint
representation of $U(N)$ will be denoted by $R$ hence the
state $\mid s \rangle$ is in the representation $
Sym(\otimes_{i=1}^{n_B} R) \times Alt(\otimes_{i=1}^{n_F} R)$ where
$Alt$ and $Sym$ stand for the anti-symmetrization and
symmetrization of the tensor product respectively. Let us denote the
number of $U(N)$ singlets in $(n_B,n_F)$ sector by $D^{U(N)}_{n_B,n_F}$. Using the orthogonality
of group characters, we have

\[
D^{U(N)}_{n_B,n_F}= \int
d \mu_{U(N)} \chi_{Sym}^{\{ n_B \}}(R)\chi_{Alt}^{[a_F]}(R),
\]
with the group invariant measure $d \mu_{U(N)}$. The symmetric and
antisymmetric powers of $R$, $\chi_{Sym}^{\{ n_B \}}(R)$,
$\chi_{Alt}^{[n_F]}(R)$ and the characters $\chi$ can be readily
constructed using the Frobenius formula (  the
complete calculation is given in the Appendix A ).

The direct evaluation of $D^{U(N)}_{n_B,n_F}$ is difficult, however we can
evaluate it indirectly introducing the following generating
functions

\begin{equation}
G^{U(N)}(a,b)=\sum_{n_B=0}^{\infty}\sum_{n_F=0}^{\infty}
D^{U(N)}_{n_B \ n_F }a^{n_B}(-b)^{n_F} , \ \ \ \ \mid a \mid <1, \ \ \ \ b\in \mathbb{R}. \label{7}
\end{equation}
The sum over $n_F$ is in fact finite ( due to the Pauli
principle, i.e.  $\chi_{Alt}^{[k]}(R)=0$ for $k>N^2$ for $U(N)$ ) but
it is more convenient to work with infinite sum. In Appendix A we
derive the following integral representation

\begin{equation}
G^{U(N)}(a,b) =\left(\frac{1-b}{1-a}\right)^N
\int_0^{2\pi}\prod_i\frac{d\alpha_i}{2\pi} \prod_{i \ne
j}\left(1-\frac{z_i}{z_j} \right)\frac{1-b\frac{z_i}{z_j}}{1-a\frac{z_i}{z_j}}, \ \ \ \ z_j=e^{i\alpha_j}. \label{8}
\end{equation}
A similar expression was derived by Skagerstam \cite{Ska} in the
context of a singlet ideal gas in flat space.

Above integrals can be calculated explicitly for $b=0$. We have ( see Appendix B )

\begin{equation}
G^{U(N)}(a,0)=\prod_{k=1}^N\frac{1}{(1-a^k)}=\sum_{n_B=0}^{\infty} a^{n_B}p_N(n_B), \label{9}
\end{equation}
where $p_N(n_B)$ is the number of partitions of $n_B$ into numbers
$1,2\ldots,N$, i.e. the number of natural solutions of the equation $\sum_{k=1}^N ki_k=n_B$.

In short the calculation goes as follows. First we
use the Cauchy determinant formula and rewrite the expression under
the integral (\ref{8}) as a determinant of a certain matrix. Then we
express the determinant as the sum over cycles and integrate each
cycle separately. Finally, we use the Cayley identity to write the result in a compact form.

 According to Eqn. (\ref{8}), $G^{U(N)}(a,b)$ is a polynomial in variable $b$.
The  coefficient next to $b^0$ is equal
$G^{U(N)}(a,0)$. We now write (\ref{8}) as
\begin{equation}
G^{U(N)}(a,b)=\left( \prod_{k=1}^N\frac{1}{(1-a^k)} \right)\sum_{i=0}^{N^2} (-1)^ib^ic^{U(N)}_i(a), \label{10}
\end{equation}
where $c^{U(N)}_i(a)$ are polynomials in variable $a$ and $c^{U(N)}_0(a)=1$.
\footnote{At this stage it is not evident that  one can factor out
the term $G^{U(N)}(a,0)$ from $G^{U(N)}(a,b)$ leaving $c^{U(N)}_i(a)$'s in the form of
polynomials. However it is indeed the case as we shall see in the
next subsection.}

The case of $SU(N)$ group is analogous. We have ( see Appendix B )
\[
G^{SU(N)}(a,b)=\left( \prod_{k=2}^N\frac{1}{(1-a^k)} \right)\sum_{i=0}^{N^2-1} (-1)^ib^ic^{SU(N)}_i(a),
\]
\[
G^{SU(N)}(a,0)= \prod_{k=2}^N\frac{1}{(1-a^k)}=\sum_{n_B=0}^{\infty} a^{n_B}q_N(n_B),
\]
where $q_N(n_B)$ is the number of partitions of $n_B$ into numbers
$2,3,\ldots,N$ and $c^{SU(N)}_i(a)$ are polynomials in variable $a$.

The determination of $c^{U(N)}_i(a)$ or $c^{SU(N)}_i(a)$ is difficult for
arbitrary $N$, however we can can get an idea about their structure by
simply counting the constructed gauge singlets.

\subsection{Explicit construction of gauge invariant states }

The formula (\ref{9}) can be derived by directly counting  the $U(N)$ singlets. If we introduce the matrices
$a^{\dagger}={a^{\dagger}}^bT_b$ where $T_b$ are $U(N)$ generators \footnote{We use the
following conventions for $U(N)$ and $SU(N)$ generators
\[
T_aT_b=\frac{1}{2N}\delta_{ab}\mathbf{1}+(d_{abc}+if_{abc})T_c,
\]
where $d_{abc}$ and $f_{abc}$ are corresponding structure tensors.}
then all  the singlets are
linear combination of the following states

\begin{equation}
\mid i_1,i_2,\ldots, i_N \rangle =(a^{\dagger})^{i_1}
({a^{\dagger}}^2)^{i_2}\ldots
({a^{\dagger}}^N)^{i_N}\mid 0 \rangle, \label{11}
\end{equation}
where we used the shorthand notation $(A)=Tr(A)$ where $A$ is an
arbitrary matrix. Higher powers of ${a^{\dagger}}$ do not appear due
to the Cayley-Hamilton theorem. Vectors (\ref{11}) are already
linearly independent therefore they form a basis in the space of
$U(N)$ singlets. For a given number of quanta $n_B$ the number of
vectors (\ref{11}) is equal to the number of natural solutions of
the equation $\sum_{k=1}^N ki_k=n_B$, i.e. there are $p_N(n_B)$ such
vectors. Therefore the generating function is exactly  (\ref{9}).

The advantage of the explicit integration over characters is that we
can compute the number of singlets in all fermionic sectors. The direct construction of such
states is possible introducing the matrices
${f^{\dagger}}={f^{\dagger}}^bT_b$. It follows that all the singlets
can be obtained by acting with $({a^{\dagger}}^i{f^{\dagger}}^j
\ldots)$ operators on  $\mid 0 \rangle$. There are however new
identities among such vectors that make many of them linearly
dependent. The process of choosing the independent ones is in
general tedious therefore the polynomials $c^{U(N)}_i(a)$ are
difficult to obtain by simply counting the constructed singlets. On the
contrary, using the character method this is straightforward ( for given $N$).

From (\ref{8}) we see that
\[
G^{U(N)}(a,b)=b^{N^2}G^{U(N)}(a,1/b),
\]
therefore
\[
c^{U(N)}_i(a)=c^{U(N)}_{N^2-i}(a),
\]
hence
\begin{equation}
D^{U(N)}_{n_B,n_F}=D^{U(N)}_{n_B,N^2-n_F}. \label{12}
\end{equation}
Equation (\ref{12}) is the simplest example of  constraints on
$D^{U(N)}_{n_B,n_F}$. It follows that the number  of gauge invariant
states in $(n_B,n_F)$ sector  is equal to the number of gauge
invariant states in $(n_B,N^2-n_F)$ sector. The identity is related
to the particle-hole symmetry, i.e. the invariance of the system under the
transformation $f^{\dagger \ a} \to f^a$, $f^a \to f^{\dagger \ a}$
(the empty fermionic states are replaced by filled fermionic ones).

 Examples of $c^{U(N)}_i(a)$ and
$c^{SU(N)}_i(a)$ for $N=2,3,4$ are in the Appendix C. We can determine these
polynomials for low values of
$N$ by explicitly constructing the singlet states. Let us start
with the case of $SU(2)$. According to character integrals we have
( see Appendix C )
\[
c^{SU(2)}_0=1, \ \ \ \ c^{SU(2)}_1=a, \ \ \ \ c^{SU(2)}_2=a, \ \ \ \ c^{SU(2)}_3=1,
\]
and the corresponding states are
\[
\mid i \rangle=({a^{\dagger}}^2)^i\mid 0 \rangle, \ \ \ \
(a^{\dagger}f^{\dagger})\mid i \rangle, \ \ \ \
(a^{\dagger}f^{\dagger}f^{\dagger})\mid i \rangle, \ \ \ \
(f^{\dagger}f^{\dagger}f^{\dagger})\mid i \rangle.
\]
There are no other, independent singlet states which (for $SU(2)$
group) can  be seen from the following simple argument. The $SU(2)$
tensor $T_{a \ldots b \ldots}$ in (\ref{T}) can only be made out of
linear combinations of products and contractions between the $SU(2)$
primite tensors $\epsilon_{abc}$ and $\delta_{ab}$. However products
( and in particular contractions) of $\epsilon_{abc}$'s can be
expressed as linear combinations of products of $\delta_{ab}$'s.
Therefore, $T_{a\ldots b \ldots}$ is either a product of
$\delta_{ab}$'s or a product of $\delta_{ab}$'s and one
$\epsilon_{abc}$. Since for $SU(2)$ group $\epsilon_{abc} \propto
(T_aT_bT_c)$ we arrive at the following states
\[
\mid i \rangle=({a^{\dagger}}^2)^i\mid 0 \rangle, \ \ \ \
(a^{\dagger}f^{\dagger})\mid i \rangle, \ \ \ \
(a^{\dagger}f^{\dagger}f^{\dagger})\mid i \rangle, \ \ \ \
(f^{\dagger}f^{\dagger}f^{\dagger})\mid i \rangle, \ \ \ \ (a^{\dagger}f^{\dagger})(a^{\dagger}f^{\dagger}f^{\dagger})\mid i \rangle,
\]
(according to the fermion exclusion principle we have $(f^{\dagger}f^{\dagger})=0$ and $(a^{\dagger}f^{\dagger})^2=0$ hence there are no operators of this form).
The last state is in fact proportional to the last but one. To see this we write
\[
(a^{\dagger}f^{\dagger})(a^{\dagger}f^{\dagger}f^{\dagger}) \propto 
a^{\dagger}_1f^{\dagger}_1  \
a^{\dagger}_1f^{\dagger}_2f^{\dagger}_3 +
a^{\dagger}_2f^{\dagger}_2 \
a^{\dagger}_2f^{\dagger}_3f^{\dagger}_1 +
a^{\dagger}_3f^{\dagger}_3 \
a^{\dagger}_3f^{\dagger}_1f^{\dagger}_2 \propto
(a^{\dagger}a^{\dagger})(f^{\dagger}f^{\dagger}f^{\dagger})
\]

The above arguments are difficult to generalize for $SU(N>2)$ due to
the additional completely symmetric tensors $d_{abc}$ that have to
be considered and the fact that the products of $f_{abc}$ tensors
cannot be expressed in terms of products of $\delta_{ab}$'s. It is
more convenient use another method, which takes advantage of the
Cayley-Hamilton theorem and which can be generalized for $SU(N>2)$.
In section 2.1.1. we give the details of this approach for the
$SU(3)$ group.

The case of $U(2)$ is more complicated.  We have
\[
c^{U(2)}_0=1, \ \ \ \ c^{U(2)}_1=1+a, \ \ \ \ c^{U(2)}_2=2a, \ \ \ \ c^{U(2)}_3=1+a,\ \ \ \ c^{U(2)}_4=1,
\]
and the corresponding singlets are
\[
\mid i,j \rangle= (a^{\dagger})^i
({a^{\dagger}}^2)^j\mid 0 \rangle \longleftrightarrow
c^{U(2)}_0=1,
\]
\[
(f^{\dagger})\mid i,j \rangle, \ \ \ \
(a^{\dagger}f^{\dagger})\mid i,j \rangle \longleftrightarrow
c^{U(2)}_1=1+a,
\]
\[
(f^{\dagger})(a^{\dagger}f^{\dagger})\mid i,j \rangle, \ \ \ \
(a^{\dagger}f^{\dagger}f^{\dagger})\mid i,j \rangle
\longleftrightarrow c^{U(2)}_2=2a,
\]
\[
(f^{\dagger}f^{\dagger}f^{\dagger})\mid i,j \rangle, \ \ \ \
(a^{\dagger}f^{\dagger}f^{\dagger})(f^{\dagger})\mid i,j \rangle
\longleftrightarrow c^{U(2)}_3=1+a,
\]
\[
(f^{\dagger})(f^{\dagger}f^{\dagger}f^{\dagger})\mid i,j \rangle \longleftrightarrow c^{U(2)}_4=1.
\]
The only difference between the $SU(2)$ case are the additional
operators $(a^{\dagger})$ and $(f^{\dagger})$ that have to be
considered when constructing the singlet state.

The construction of independent  states for $SU(3)$ or $U(3)$, in
all fermion sectors, is already very nontrivial  and will be
discussed in section 2.1.1. However, for $n_F=0,1$ we can construct
them for arbitrary $U(N)$ and $SU(N)$. They are
\[
\mid i_1,i_2,\ldots, i_N \rangle \ \  \longleftrightarrow \ \ c^{U(N)}_0=1,
\]
\[
(f^{\dagger}{a^{\dagger}}^k)\mid i_1,i_2,\ldots, i_N \rangle,  \ k=0,\ldots,N-1 \ \
\longleftrightarrow \ \  c^{U(N)}_1=1+a+\ldots+a^{N-1},
\]

\[
\mid 0,i_2,\ldots, i_N \rangle \ \  \longleftrightarrow  \ \ c^{SU(N)}_0=1,
\]
\[
(f^{\dagger}{a^{\dagger}}^k)\mid 0,i_2,\ldots, i_N \rangle,  \ k=1,\ldots,N-1 \ \
\longleftrightarrow  \ \ c^{SU(N)}_1=a+\ldots+a^{N-1}.
\]

From the explicit results  for $N=2,3,4$ we see that there is a relation
between the generating functions for $U(N)$ and $SU(N)$ namely

\[
G^{U(N)}(a,b)=\frac{1-b}{1-a}G^{SU(N)}(a,b).
\]
It can be understood in terms of gauge singlets just constructed. For
the case of $U(N)$ we have additional trace operators $(a^{\dagger})$ and
$(f^{\dagger})$ ( compared to the $SU(N)$ case) therefore to obtain all the
$U(N)$ singlets we have to multiply the $SU(N)$ singlets by
$(a^{\dagger})^k$, $k \ge 0$ and by  $(f^{\dagger})^k$, $k=0,1$. In terms of
generating functions this corresponds to multiplying
$G^{SU(N)}(a,b)$ by $\frac{1-b}{1-a}$.

\subsubsection{The construction of $SU(3)$ invariant states}

The explicit evaluation of the integral (\ref{8}) for $N=3$ gives

\[
c^{SU(3)}_0=1, \ \ \ \ c^{SU(3)}_1=a+a^2, \ \ \ \ c^{SU(3)}_2=a+a^2+2a^3,
\]
\[
c^{SU(3)}_3=1+a+2a^2+3a^3+a^4, \ \ \ \ c^{SU(3)}_4=2a+4a^2+2a^3+2a^4, \ \ \ \ c^{SU(3)}_i=c^{SU(3)}_{8-i}.
\]
It follows that $D^{SU(3)}_{n_B,n_F}$ with $n_F>0$ can be obtained
by linear combinations of $D^{SU(3)}_{n_B,0}=q_3(n_B)$. According to the above
results we have
\[
D^{SU(3)}_{n_B,0}=q_3(n_B),
\]
\[
D^{SU(3)}_{n_B,1}=q_3(n_B-1)+q_3(n_B-2),
\]
\[
D^{SU(3)}_{n_B,2}=q_3(n_B-1)+q_3(n_B-2)+2q_3(n_B-3),
\]
\[
D^{SU(3)}_{n_B,3}=q_3(n_B)+q_3(n_B-1)+2q_3(n_B-2)+3q_3(n_B-3)+q_3(n_B-4),
\]
\begin{equation}
D^{SU(3)}_{n_B,4}=2q_3(n_B-1)+4q_3(n_B-2)+2q_3(n_B-3)+2q_3(n_B-4). \label{20}
\end{equation}
We now construct the states in Fock space corresponding to these numbers.
The construction is done separately in sectors with given number of fermionic quanta.

The bases  for
$n_F=0$ and  $n_F=1$ sectors are

\begin{equation}
\mid i,j \rangle = (a^{\dagger}a^{\dagger})^i (a^{\dagger}a^{\dagger}a^{\dagger})^j\mid 0 \rangle \longleftrightarrow 1, \label{21}
\end{equation}
and
\[
(a^{\dagger}f^{\dagger})\mid i,j \rangle \longleftrightarrow a,
\]
\begin{equation}
(a^{\dagger}a^{\dagger}f^{\dagger})\mid i,j \rangle \longleftrightarrow  a^2. \label{22}
\end{equation}
The number of vectors (\ref{21}) with given number of quanta
$n_B$ is exactly $q_3(n_B)$ and the number of vectors (\ref{22}) with given
number of quanta  $n_B$ is precisely $q_3(n_B-1)+ q_3(n_B-2)$, i.e. the
number of states $(a^{\dagger}f^{\dagger})\mid i,j \rangle$ with
$n_B$ bosons is $q_3(n_B-1)$ and the number of states
$(a^{\dagger}f^{\dagger}f^{\dagger})\mid i,j \rangle$ with $n_B$
bosons is $q_3(n_B-2)$. This result agrees with $D^{SU(3)}_{n_B,0}$ and
$D^{SU(3)}_{n_B,1}$ in (\ref{20}).

 Our strategy to determine the basis in sectors with
$n_F>1$ is the following. First we list all possible trace operators
with given number of fermions. Then, with use of the Cayley-Hamilton
theorem and some symmetry arguments, we choose only the linearly
independent ones. The resulting trace operators acting on states
$\mid i,j \rangle$ are our candidates for the basis. In order to
find out whether they really form a basis we compare the number of
singlets constructed in this way with $D^{SU(3)}_{n_B,n_F}$ in (\ref{20})
(or equivalently associate the corresponding polynomial $c^{SU(3)}_i(a)$).

In $n_F=2$ sector the possible trace operators are
\[
(a^{\dagger}f^{\dagger}f^{\dagger}), \ \ \ \ \   n_B=1,
\]
\[
(a^{\dagger}a^{\dagger}f^{\dagger}f^{\dagger}), \  
(a^{\dagger}f^{\dagger}a^{\dagger}f^{\dagger})=0, \  
(a^{\dagger}f^{\dagger})(a^{\dagger}f^{\dagger})=0, \ \ \ \ n_B=2,
\]
\[
(a^{\dagger}a^{\dagger}f^{\dagger}a^{\dagger}f^{\dagger}), \  
(a^{\dagger}f^{\dagger})(a^{\dagger}a^{\dagger}f^{\dagger}), \ \ \
\ n_B=3,
\]
\[
(a^{\dagger}a^{\dagger}f^{\dagger}a^{\dagger}a^{\dagger}f^{\dagger})=0, \ 
(a^{\dagger}a^{\dagger}f^{\dagger})(a^{\dagger}a^{\dagger}f^{\dagger})=0,
\ \ \ \ n_B=4.
\]
The other trace operators involve at least one ${a^{\dagger}}^3$ hence they are linearly dependent.
We have already taken into consideration the cyclicity
of the trace. Therefore, there are four families of $SU(3)$
invariant states in this sector
\[
(a^{\dagger}f^{\dagger}f^{\dagger})\mid i,j \rangle  \longleftrightarrow a
\]
\[
(a^{\dagger}a^{\dagger}f^{\dagger}f^{\dagger})\mid i,j \rangle  \longleftrightarrow a^2
\]
\begin{equation}
(a^{\dagger}a^{\dagger}f^{\dagger}a^{\dagger}f^{\dagger})\mid i,j \rangle, \ 
(a^{\dagger}f^{\dagger})(a^{\dagger}a^{\dagger}f^{\dagger})\mid i,j \rangle  \longleftrightarrow  2a^3. \label{23}
\end{equation}

 In the sector with $n_F=3$  the possible trace operators are
\[
({f^{\dagger}}^3), \ \ \ \  n_B=0,
\]
\[
(a^{\dagger}{f^{\dagger}}^3),  \ \ \ \ n_B=1,
\]
\[
({a^{\dagger}}^2{f^{\dagger}}^3), \ 
(a^{\dagger}f^{\dagger}a^{\dagger}{f^{\dagger}}^2), \ 
(a^{\dagger}f^{\dagger})(a^{\dagger}{f^{\dagger}}^2),  \ \ \ \
n_B=2,
\]
\[
({a^{\dagger}}^2f^{\dagger}a^{\dagger}{f^{\dagger}}^2), \ 
({a^{\dagger}}^2{f^{\dagger}}^2a^{\dagger}f^{\dagger}), \ 
(a^{\dagger}f^{\dagger}a^{\dagger}f^{\dagger}a^{\dagger}f^{\dagger}), \ 
(a^{\dagger}f^{\dagger})({a^{\dagger}}^2{f^{\dagger}}^2), \ 
({a^{\dagger}}^2f^{\dagger})(a^{\dagger}{f^{\dagger}}^2), \ \ \ \
n_B=3,
\]
\[
({a^{\dagger}}^2f^{\dagger}{a^{\dagger}}^2{f^{\dagger}}^2), \ 
({a^{\dagger}}^2f^{\dagger}a^{\dagger}f^{\dagger}a^{\dagger}f^{\dagger}), \ 
(a^{\dagger}f^{\dagger})({a^{\dagger}}^2f^{\dagger}a^{\dagger}f^{\dagger}), \ 
(a^{\dagger}a^{\dagger}f^{\dagger})(a^{\dagger}a^{\dagger}f^{\dagger}f^{\dagger}),\
\ \ \ n_B=4,
\]
\[
({a^{\dagger}}^2f^{\dagger}{a^{\dagger}}^2f^{\dagger}a^{\dagger}f^{\dagger}), \ 
({a^{\dagger}}^2f^{\dagger})({a^{\dagger}}^2f^{\dagger}a^{\dagger}f^{\dagger}),\
\ \ \ n_B=5,
\]
\[
({a^{\dagger}}^2f^{\dagger}{a^{\dagger}}^2f^{\dagger}{a^{\dagger}}^2f^{\dagger}), \ \ \ \ n_B=6.
\]
Again, other trace operators involve at least one ${a^{\dagger}}^3$
hence they are linearly dependent. From the 17 operators listed
above only 8 are linearly independent. The linear dependence is due
to the identity for $SU(3)$ generators \footnote{  The complete symmetrization over indices is
\[
T_{(a}T_bT_{c)}=T_aT_bT_c+permutations,
\]
without the conventional $\frac{1}{3!}$ factor.
}
\[
T_{(a}T_bT_{c)}=\delta_{(ab}T_{c)}+4 d_{abc}\mathbf{1}.
\]
 This identity is
equivalent to the Cayley-Hamilton theorem for $3 \times 3$ traceless
matrices. If we contract its left and right hand side with
${a^{\dagger}}_a {a^{\dagger}}_b {f^{\dagger}}_c$ we will obtain an
analogue of Cayley-Hamilton theorem (with fermionic matrices)
\[
a^{\dagger}a^{\dagger}f^{\dagger}+
f^{\dagger}a^{\dagger}a^{\dagger}+a^{\dagger}f^{\dagger}a^{\dagger}=
\frac{1}{2}(a^{\dagger}a^{\dagger})f^{\dagger}+
(a^{\dagger}f^{\dagger})a^{\dagger}+(a^{\dagger}a^{\dagger}f^{\dagger}).
\]
All the linearly dependent trace operators can be derived from the
last identity ( see Appendix D ) and one finds that there are only 8
families of independent vectors. They are
\[
({f^{\dagger}}^3)\mid i,j \rangle \longleftrightarrow 1,
\]
\[
(a^{\dagger}{f^{\dagger}}^3)\mid i,j \rangle \longleftrightarrow a,
\]
\[
(a^{\dagger}a^{\dagger}{f^{\dagger}}^3)\mid i,j \rangle, \ \ \ \
(a^{\dagger}f^{\dagger})(a^{\dagger}{f^{\dagger}}^2)\mid i,j
\rangle \longleftrightarrow 2a^2,
\]
\[
({a^{\dagger}}^2{f^{\dagger}}^2a^{\dagger}f^{\dagger})\mid i,j
\rangle, \ \ \ \
({a^{\dagger}}^2f^{\dagger})(a^{\dagger}{f^{\dagger}}^2)\mid i,j
\rangle, \ \ \ \
(a^{\dagger}f^{\dagger})({a^{\dagger}}^2{f^{\dagger}}^2)\mid i,j
\rangle  \longleftrightarrow 3a^3,
\]
\begin{equation}
({a^{\dagger}}^2f^{\dagger})({a^{\dagger}}^2{f^{\dagger}}^2)\mid i,j \rangle \longleftrightarrow a^4. \label{24}
\end{equation}

The $n_F=4$ sector is the most complicated one. There  are 52
trace operators that one can construct  however only 10 of them are
independent.  They are
\[
(a^{\dagger}{f^{\dagger}}^3)\mid i,j \rangle, \ \ \ \
(a^{\dagger}f^{\dagger})({f^{\dagger}}^3)\mid i,j \rangle \longleftrightarrow 2a,
\]
\[
({a^{\dagger}}^2{f^{\dagger}}^4)\mid i,j \rangle, \ \ \ \
({a^{\dagger}}^2f^{\dagger})({f^{\dagger}}^3)\mid i,j \rangle, \ \
\ \ (a^{\dagger}f^{\dagger})(a^{\dagger}{f^{\dagger}}^3)\mid i,j
\rangle, \ \ \ \
(a^{\dagger}{f^{\dagger}}^2)(a^{\dagger}{f^{\dagger}}^2)\mid i,j
\rangle \longleftrightarrow 4a^2,
\]
\[
({a^{\dagger}}^2f^{\dagger})(a^{\dagger}{f^{\dagger}}^3)\mid i,j
\rangle, \ \ \ \
(a^{\dagger}{f^{\dagger}}^2)({a^{\dagger}}^2{f^{\dagger}}^2)\mid
i,j \rangle  \longleftrightarrow 2a^3,
\]
\[
(a^{\dagger}f^{\dagger})({a^{\dagger}}^2f^{\dagger})(a^{\dagger}{f^{\dagger}}^2)\mid
i,j \rangle, \ \ \ \
(a^{\dagger}{f^{\dagger}}^2)({a^{\dagger}}^2f^{\dagger}a^{\dagger}f^{\dagger})\mid
i,j \rangle  \longleftrightarrow 2a^4.
\]
The method to extract these 10  independent ones is the same as
in the $n_F=3$ case.

\subsection{Physical constraints on $D^{U(N)}_{n_B,n_F}$}
There is a class of identities analogous
to (\ref{12}) which also have a physical interpretation.
For example if we put $b=1$ in (\ref{8}) we obtain
\[
G^{U(N)}(a,1)=0,
\]
therefore
\begin{equation}
\sum_{n_F \ even}D^{U(N)}_{n_B,n_F}=\sum_{n_F \ odd} \label{13}
D^{U(N)}_{n_B,n_F}.
\end{equation}
The above equation is suggestive of supersymmetry since the number
of fermionic states ( the sum over odd $n_F$ ) equals to the number
of bosonic degrees of freedom  (the sum over even $n_F$ ). However,
since we did not specify any hamiltonian, Eq. (\ref{13}) tells us
that the bosonic and fermionic states match, as they should, i.e.
the Hilbert space of singlets is already "prepared" for
supersymmetry.

An interesting observation is given by equation (\ref{7}), when $a=b$. The result is
of a form of the Witten index. Indeed, if we consider supersymmetric
harmonic oscillator
\[
H=\{Q,Q^{\dagger}\}=a^{\dagger \ b}a^b+f^{\dagger \ b}f^b, \ \ \ \ Q=a^{\dagger \ b}f^b,
\]
then the energy is proportional to the number of quanta hence
there is only one vacuum state and naturally it is the Fock vacuum. Therefore, the Witten index is 1.
We confirm that by explicitly computing $G^{U(N)}(a,a)$ using Eqn. (\ref{8})
from which the identity $G^{U(N)}(a,a)=1$ follows.
 We also note that if we put $b=a$ in (\ref{7}) we obtain
\[
G^{U(N)}(a,a)=\sum_{n_B, \ n_F}
(-1)^{n_F}D^{U(N)}_{n_B,n_F}a^{n_B+n_F}=1+\sum_{k>0}\sum_{n_B+n_F=k}
(-1)^{n_F}D^{U(N)}_{n_B,n_F}a^{n_B+n_F},
\]
therefore
\[
\sum_{n_B+n_F=\textit{const.}>0} (-1)^{n_F}D^{U(N)}_{n_B,n_F}=0.
\]
The origin of the above identities lies in the dynamics of the
supersymmetric harmonic oscillator although it is perhaps not
evident at first sight.

Another example of a hamiltonian which brings physical meaning to
some identities including $D^{U(N)}_{n_B,n_F}$'s is the hamiltonian given
by the supercharge \cite{VW}
\[
Q=f^ba^{\dagger \ b}+gd_{abc}a^{\dagger \ a}
a^{\dagger \ b}f^c.
\]
One can show \cite{WVspin} that the in the limit of strong 't Hooft
coupling $\lambda = Ng^2 \rightarrow \infty$  the energies
are  proportional to $n_B+2n_F$ and that the supercharges act in
the subspace of vectors such that $n_B+2n_F$ is fixed. Therefore
the contribution to the Witten index in terms of generating
function is now $G^{U(N)}(a,a^2)$. This quantity is not
necessarily a constant in variable $a$ since there may be other vacua in fermion
sectors. However there is a finite number of vacua hence
$G^{U(N)}(a,a^2)$ is at most a polynomial in $a$. We confirm
this by explicitly calculating $G^{U(N)}(a,a^2)$ and $G^{SU(N)}(a,a^2)$ for the
lowest values of $N$. We have

\[
G^{U(2)}(a,a^2)=1+a+a^2+a^5, \ \ \ \
G^{U(3)}(a,a^2)=1+a+a^2+a^3+a^5+a^6+a^7-a^9+a^{10}+a^{11},
\]
and
\[
G^{SU(2)}(a,a^2)=1+a^2-a^3+a^4, \ \ \ \
G^{SU(3)}(a,a^2)=1+a^2+a^5+a^7 -a^8 + a^{10}.
\]
The coefficients of polynomials $G^{H}(a,a^2)$ where $H=U(N),SU(N)$  give us the
difference between the number of bosonic and fermionic vacua. In
general, since $G^{H}(a,a^2)$ is a polynomial, the
constraint for $D^{H}_{n_B,n_F}$'s, coming from Eqn. (\ref{7}),
is now
\[
\sum_{n_B+2n_F=k}(-1)^{n_F} D^{\textbf{H}}_{n_B,n_F}=0,
\]
for $k$ greater then some $k_0$. It also seems that
\[
 \sum_{n_B+2n_F=k}(-1)^{n_F} D^{\textbf{H}}_{n_B,n_F}=0,1,-1,
\]
for $k \le k_0$.

We find it very interesting that although $D^{H}_{n_B,n_F}$'s are
just some group theory numbers, they are constrained by the
dynamics of the properly chosen supersymmetric hamiltonian.

\section{The character method for $D=3+1,5+1,9+1$ SYMQM}

The generalization of the $D=2$ case to
$D=4,6,10$ cases is now straightforward.
 The bosonic and
fermionic creation operators ${a^{\dagger}}_i^{b}$,
${f^{\dagger}}_{\alpha}^{b}$ are now labeled by color index $b$,
spatial index $i=1,\ldots,d$ and spinor index
\[
\alpha=1,2 \ \ \ \mbox{for D=3+1 }, \ \ \ \
\alpha=1,2,3,4 \ \ \ \mbox{for D=5+1 }, \ \ \ \
 \alpha=1,\ldots,8 \ \ \ \mbox{for D=9+1 }.
\]
The state with $n_B$ bosons and $n_F$ fermions is now
\[
{a^{\dagger}}_{i_1}^{b_1} \ldots {a^{\dagger}}_{i_{n_B}}^{b_{n_B}} {f^{\dagger}}_{{\alpha}_1}^{c_1}
\ldots {f^{\dagger}}_{{\alpha}_{n_F}}^{c_{n_F}}\mid 0 \rangle,
\]
and the number of $U(N)$ singlets is

\[
D^{ U(N), d}_{n_B  \ n_F}
= \int d\mu_{\textbf{U(N)}}\chi_{Sym}(R_B^{n_B}) \chi_{Alt}(R_F^{n_F}),
\]
where
\[
\chi(R_B)=d \ \chi^{d=1}(R), \ \ \ \ \chi(R_F)=(d-1)\chi^{d=1}(R).
\]
The above equations for $\chi(R_B)$  and $\chi(R_F)$ are in fact
the only difference between the $d=1$ case. We can introduce the
generating functions analogous to (\ref{7}) and perform the same
manipulations to find that the corresponding generating functions
are

\begin{equation}
G^{ U(N), d}(a,b) =\frac{(1-b)^{N(d-1)}}{(1-a)^{Nd}}
\int_0^{2\pi} \prod_i\frac{d\alpha_i}{2\pi} \prod_{i \ne
j}(1-\frac{z_i}{z_j})\frac{(1-b\frac{z_i}{z_j})^{d-1}}{(1-a\frac{z_i}{z_j})^d}. \label{14}
\end{equation}
From (\ref{14}) we identify the particle-hole symmetry
\[
D^{ U(N), d}_{n_B,n_F}=D^{ U(N), d}_{n_B,N^2(d-1)-n_F},  \ \ \ \
\]
and supersymmetry
\[
\sum_{n_F \ even}D^{ U(N), d}_{n_B,n_F}= \sum_{n_F \ odd}D^{ U(N), d}_{n_B,n_F}.
\]
Taking $b=a$ and using the results from previous section we obtain
\[
G^{ U(N), d }(a,a) =\frac{1}{\prod_{k=1}^N(1-a^k)},
\]
therefore
\[
\sum_{n_F+n_B=k}(-1)^{n_F}D^{ U(N), d }_{n_B,n_F}=p_N(k).
\]
The result does not depend on $d$ which is surprising but possible
since $G^{ U(N), d }(a,a)$ is the generating function for the
differences between bosonic and fermionic gauge invariant states. However, $G^{ U(N), d}(a,a)$ cannot be
interpreted as the contribution to the Witten index for
supersymmetric harmonic  oscillator in $d+1$ dimensions. This is
because the supersymmetric harmonic oscillator with gauge degrees of
freedom does not exist in $d+1>2$ dimensions together with the
singlet constraint. Such system cannot exist since the number of
bosonic and fermionic degrees of freedom do not mach. For example
there are $d$ states with $n_B=1$ and $n_F=0$, they are
$(a^{\dagger}_i)$, $i=1,\ldots, d$. On the other hand there are
$d-1$ states with $n_B=0$ and $n_F=1$, they are
$(f^{\dagger}_{\alpha})$, $\alpha=1,\ldots, d-1$. The difference
is precisely equal $p_N(1)=1$.

The generating function can be computed explicitly for  arbitrary value of $N$
although the general $N$ dependence  is difficult to obtain even
for $b=0$. The case $N=2$ is particularly easy to evaluate, we have
\begin{equation}
G^{ SU(2), d}(a,0)=-\frac{1}{2}\int \frac{dz_1}{2\pi i} z_1^{d-2}
\frac{(z_1-1)^2}{(z_1-a)^d(1-az_1)^d}=-\frac{1}{2(d-1)!}\frac{d^d}{dz^d}\left(
\frac{z^{d-2}(z-1)^2}{(1-az)^d} \right)\mid_{z=a}. \label{15}
\end{equation}
 The cases with
$N=3,4$, $d=3,5,9$ are presented in Appendix E while the
values of $D^{ SU(2), 9}_{n_B,n_F}$ for $n_F\le 12$, $n_B \le 10$ are presented in Tables 1, 2 and 3.

   \begin{table}[tbp]
    \begin{tabular}{ccccccccc} \hline\hline
$n_F$ & $ 0 $ & $ 1 $ & $ 2 $ & $ 3 $  & $ 4 $\\

$n_B$ & $   $ & $   $ & $   $ & $   $  &    \\
    \hline
   $0$  & 1 & 72 & 28 & 120 & 406   \\
   $1$  & 0 & 288 & 324 & 2016 & 9072 \\
   $2$  & 45 & 3240 & 3816 & 21024 & 89838  \\
   $3$  & 84 & 12960 & 23652 & 150360 & 692874   \\
   $4$  & 1035 & 74520 & 144000 & 882720 & 4049640  \\
   $5$  & 2772 & 270864 & 662436 & 4331880 & 20528802 \\
   $6$  & 16215 & 1119096 & 2906448 & 18805104 & 89459160  \\
   $7$  & 46530 & 3635280 & 10912572 & 72993096 & 353298330 \\
   $8$  & 189288 & 12260160 & 38914524 & 259803720 & 1263689658  \\
 \hline\hline
    \end{tabular}
\caption{Number of $SU(2)$ singlets for $D=10$ spacetime dimensions
in sectors with  $ 0 \leq n_F \leq 4$ and $0 \leq n_B \leq 8$.}
 \end{table}

   \begin{table}[tbp]
    \begin{tabular}{ccccccccccccccc} \hline\hline
$n_F$ & $ 5 $ & $ 6 $ & $ 7 $ & $ 8 $ \\

$n_B$ & $ $   & $ $   & $ $   &     \\
    \hline
   $0$   & 1512  &  4060 & 8856 & 17605\\
   $1$   & 29232 & 81648 & 192528 &  374544\\
   $2$   & 321048 & 907452 & 2121192 & 4230801\\
   $3$   & 2426928 & 6998040 & 16742544 & 33436080\\
   $4$   & 14752080 & 42942060 & 103041000 & 208064475\\
   $5$   & 74701368 & 220014900 & 533991024 & 1081967760\\
   $6$   & 331885680 & 984408096 & 2399008272 & 4891876599\\
   $7$   & 1314510120 & 3928885884 & 9640642968 & 19721394891\\
   $8$   & 4754606472 & 14285876220 & 35181176976 & 35181176976\\
 \hline\hline
    \end{tabular}
\caption{Number of $SU(2)$ singlets for $D=10$ spacetime dimensions
in sectors with  $ 5 \leq n_F \leq 8$ and $0 \leq n_B \leq 8$.}
 \end{table}

 \begin{table}[tbp]
    \begin{tabular}{ccccccccccccccc} \hline\hline
$n_F$ & $ 9 $ & $ 10 $  & $ 11 $ & $ 12 $ & $ \Sigma $  \\

$n_B$ & $ $   & $ $     & $ $    & $ $    &  \\
    \hline
$0$  & $ 29512 $ & $ 41392 $ & $ 51520 $ & $ 56056 $ & $ 211068 $  \\
$1$  & $ 626040 $ & $ 908460 $ & $ 1126944 $ & $ 1205568 $ & $ 4556448 $\\
$2$  & $ 7158600 $ & $ 10328580 $ & $ 12886776 $ & $ 13896792 $ & $ 51966252 $  \\
$3$  & $ 56800008 $ & $ 82741428 $ & $ 103339320 $ & $ 111140484 $ & $ 414495042 $  \\
$4$  & $ 355678200 $ & $ 518416380 $ & $ 649288080 $ & $ 700074900 $ & $ 2597347530 $   \\
$5$  & $ 1856261448 $ & $ 2718108792 $ & $ 3408546960 $ & $ 3673243476 $ & $ 13592436138 $ \\
$6$  & $ 8426386704 $ & $ 12356524344 $ & $ 15522106992 $ & $ 16746545508 $ & $ 61770199986 $  \\
$7$  & 34066733976 & 50095041876 & 62995900968 & 67964640282 & 250166120589 \\
$8$  & 72270195525 & 125190973512 & 184350316788 & 232102914120 & 920270148237  \\
 \hline\hline
    \end{tabular}
\caption{Number of $SU(2)$ singlets for $D=10$ spacetime dimensions
in sectors with  $ 9 \leq n_F \leq 12$ and $0 \leq n_B \leq 8$.  $\Sigma$ gives
the cumulative size up to $n_B$.}
 \end{table}

Some of these values have been already obtained earlier \cite{WosiekD10}
with considerable numerical effort by constructing the singlets
directly . We see that the numbers of singlets grow extremely fast
in this case, e.g. $D^{ SU(2), 9}_{8,12}\approx 2.5 \cdot 10^{11}$.
It follows that the direct numerical approaches ( e.g. the cutoff
method \cite{WosiekD4} ) to SYMQM in $d=9$ dimensions is difficult
to deal with even for the fastest computers. The $d=9$ model is
particularly troublesome because the fermion number is not conserved
hence one cannot diagonalize the hamiltonian in each fermion sector
separately.

If the hamiltonian has additional $SO(d)$ symmetry then it is
convenient to work with gauge and $SO(d)$ singlets. The method
presented here can be applied also in this case by computing the
generating function for $SU(N)$ invariant states with given spin (
see Appendix F ). Fortunately all the SYMQM have the  $SO(d)$
symmetry. It is then possible that the numerical analysis of these
models is within our reach once we work in sectors with given $SO(9)$ angular
momentum.

\section{The partition functions for free Yang-Mills theories}

The generating functions $G^{U(N), d}(a,b)$ have an interesting application
in, seemingly unrelated, problem of computing the partition function
of free Yang-Mills theories on  $S^1 \times time$. We show in this
section that the partition functions of such theories can be
expressed in terms of $G^{U(N), d}(a,b)$'s in a rather simple way. Following Sundborg \cite{Sundborg}
and Aharony et al. \cite{Aharony} we write the partition function of the free Yang-Mills theory with
$n_S$, $n_V$ and $n_F$  number of scalar vector and fermion fields respectively as
\begin{equation}
Z(x)=\int d\mu_{\textbf{G}} \  exp  \left(
\sum_{m=1}^{\infty}\frac{1}{m}[Z_B(x^m)+(-1)^m Z_F(x^m)    ]\chi(R^m)
\right), \label{16}
\end{equation}
where $Z_B(x)$ and $Z_F(x)$ are bosonic and fermionic single partition functions given
explicitly, for $D=2$, by
\[
Z_B(x)=n_S z_S(x)+n_V z_V(x), \ \ \ \ Z_F(x)=n_F z_F(x),
\]
\[
z_S(x)=\frac{1+x}{1-x},  \ \
\ \ z_V(x)=x^2, \ \ \ \ z_F(x)=\frac{4\sqrt{x}}{1-x}.
\]
It is clear that $Z(x)$ is very similar to $G^{U(N),d}(a,b)$.  The only difference is that
there are no single particle partition function $z_B(x)$ and
$z_F(x)$ in $G^{U(N),d}(a,b)$, instead there are two generating parameters
$a$ and $b$.
In two dimensions scalar fields have the scaling  dimension zero
hance $z_S(0)=1$ and the partition function is divergent. To avoid
this we take $n_S=0$. Next, since $z_F(x)$ is a rational function,
we Taylor expand $z_F(x)$ and substitute it to (\ref{16}). Using the
formulas and conventions from Appendix A, (\ref{16})
becomes
\[
Z(x)= \int_{\mid z_j\mid<1} \prod_{j=1}^N
\frac{dz_j}{2\pi i z_j}\prod_{i\ne j}(1-\frac{z_i}{z_j})\prod_{k=0}^{\infty} \prod_{i,
j}\frac{(1+x^{k+\frac{1}{2}}\frac{z_i}{z_j})^{4n_F}}{(1-x^2\frac{z_i}{z_j})^{n_V}}.
\]
The above formula is more complicated then  (\ref{14}) due to the infinite
product over $k$. However, the product does not appear when we take $n_F=0$. In this case $Z(x)$ becomes

\begin{equation}
Z(x)=G^{U(N),n_V}(x^2,0).
\end{equation}
It is amusing that the partition
function of free Yang-Mills theory on $S^1 \times \mathbb{R}$ with
$n_V$ vector fields is given directly by $G^{U(N),n_V}(x^2,0)$. On
the other hand two dimensional gauge theories
have many exceptional features hence their partition
functions may resemble some simplicity. Indeed, Yang-Mills theories
on compact, orientable surfaces are exactly solvable \cite{Migdal} and their
partition functions are known to be simple expressions depending on
group theory parameters.

\section{Summary}

In this paper we focused on calculating the number of $U(N)$
($SU(N)$) singlets motivated by the numerical approach based on the
cut Fock space. Once the basis is known the cutoff method may be
used. However, the very construction of such a basis is far from
easy and proved \cite{WosiekD4,WosiekD41,WosiekD10}  to be very time
consuming when symbolic programs are involved.  The results
presented here give us an algorithm to construct such basis
theoretically thereby facilitating the numerical considerations. In
particular we hope that the results presented here applied to the
$j=0$ sector of $D=9+1$ SYMQM will help to obtain the
nonperturbative spectra of this highly complicated system.

It is interesting that the generating functions  $G^{U(N),
d}(a,b)$ have other, seemingly independent, applications, i.e.
they give rise to the Witten index in a class of models as well as
they can be used to compute the partition functions of free
Yang-Mills theories on $S^1 \times \mathbb{R}$.

\section{Acknowledgments}
I thank R. Janik, G. Veneziano and J. Wosiek for many discussions. I
also thank referees for bringing to my attention references \cite{Dolan,Ska} and for many important comments regarding this
manuscript. This work was supported by the the grant of Polish
Ministry of Science and Education no. P03B 024 27 ( 2004 - 2007 )
and N202 044 31/2444 ( 2006-2007 ) and the  Jagiellonian University
Estreicher foundation.

\section{Appendix A - Group theory conventions and the integral representation of $G^{U(N)}(a,b)$}

Here we give the conventions used in integrals over
characters and derive (\ref{8}).  They can be found in, e.g.
ref. \cite{konwencje}. The methods used in this appendix are similar to the ones used in \cite{Aharony}.

The $U(N)$ invariant measure is
\[
d\mu_{U(N)} =  \frac{1}{N!}\prod_{i=1}^{N}\frac{d\alpha_i}{2\pi} \mid M
\mid^2, \ \ \ \ d\mu_{SU(N)} =
\frac{1}{N!}\prod_{i=1}^{N}\frac{d\alpha_i}{2\pi}\delta_P(\sum_{i=1}^{N}
\alpha_i) \mid M \mid^2, \ \ \ \ \alpha_i \in [0,2\pi],
\]
where $\delta_P$ is a periodic Dirac delta with period $2\pi$
\[
\delta_P(x)=\sum_k\delta(x-2\pi k),
\]
the measure factor $M$ is given by Vandermonde determinant
\[
M=Det(z_j^{(N-i)})=\prod_{i<j} (z_i-z_j), \ \ \ \ z_j=e^{i \alpha_j},
\]
the symmetric and antisymmetric powers of $R$, $\chi_{Sym}^{\{ n_B
\}}(R)$ and $\chi_{Alt}^{[a_F]}(R)$, are given by Frobenius formulas
\[
\chi_{Sym}^{\{ n_B \}}(R)=\sum_{\sum_k k i_k=n_B} \prod_{k=1}^{n_B}
\frac{1}{i_k!}\frac{\chi^{i_k}(R^k)}{k^{i_k}},
\]

\[
\chi_{Alt}^{[a_F]}(R)=\sum_{\sum_k k i_k=n_F} (-1)^{\sum_k i_k}
\prod_{k=1}^{n_B}
\frac{1}{i_k!}\frac{\chi^{i_k}(R^k)}{k^{i_k}},
\]
and the characters $\chi$ are given by Weyl determinant formula
\[
\chi(R)\equiv\chi(\{\alpha_i\}_{i=1}^N)=
\frac{Det(z_j^{(N-i+l_i)})}{Det(z_j^{(N-i)})}, \ \ \ \ \chi(R^k)=
\chi(\{k\alpha_i\}_{i=1}^N).
\]
The numbers $l_i$ enumerate the representation in which the
character is calculated. In our case it is the adjoint
representation of $U(N)$ ( or $SU(N)$ ) therefore
$(l_1,l_2,\ldots,l_N)=(2,1,\ldots,1,0)$ . In this representation the
characters simplify into
\[
\chi_{U(N)}(\{\alpha_i\})=
 \sum_{i,j}\frac{z_i}{z_j},
\]

\[
\chi_{SU(N)}(\{\alpha_i\})=\sum_{ i,j }\frac{z_i}{z_j} -1.
\]

In order to derive (\ref{8}) let us introduce
\[
F_{Sym}(a,\{\alpha_i\}_{i=1}^N)=\sum_{n_B=0}^{\infty} a^{n_B} \chi_{Sym}(R^{n_B}), \ \ \ \ \mid a \mid <1,
\]
\[
F_{Alt}(b,\{\alpha_i\}_{i=1}^N)=\sum_{n_F=0}^{\infty} (-1)^{n_F} b^{n_F} \chi_{Alt}(R^{n_F}),
\]
\[
G^{\textbf{H}}(a,b)=\int_{[0,2\pi]^N}  d \mu_{\textbf{G}}
F_{Sym}(a,\{\alpha_i\})F_{Alt}(b,\{\alpha_i\}) ,
\]
\[
D^{\textbf{H}}_{n_B,n_F}=\frac{1}{n_B!}
\frac{(-1)^{n_F}}{n_F!}
\frac{\partial^{n_B}}{\partial
a^{n_B}}\frac{\partial^{n_F}}{\partial b^{n_F}} G^{\textbf{G}}(a,b).
\]

The last sum is in fact finite since for $U(N)$ ( or $SU(N)$ ),
$\chi_{Alt}(R^{n_F})=0$ when $n_F>N^2$ ( or $n_F>N^2-1$). It is
however more convenient to work with infinite sum as we will see in
the following. The $b$ variable is not bounded.

 Using the standard manipulations we obtain

\[
F_{Sym}(a,\{\alpha_i\}_{i=1}^N)=\exp \left( \sum_{k=1}^{\infty}
\frac{a^k}{k}\chi(\{k\alpha_i\}_{i=1}^N)\right),
\]
\[
F_{Alt}(b,\{\alpha_i\}_{i=1}^N)=\frac{1}{F_{Sym}(b,\{\alpha_i\}_{i=1}^N)}.
\]
The generating function can be calculated explicitly for arbitrary
$U(N)$ and $SU(N)$. We have

\[
F_{Sym}^{U(N)}(a,\{\alpha_i\}_{i=1}^N)=\frac{1}{\prod_{i,j}(1-a\frac{z_i}{z_j})},
\]
\[
F_{Sym}^{SU(N)}(a,\{\alpha_i\}_{i=1}^N)=(1-a)F_{Sym}^{U(N)}(a,\{\alpha_i\}_{i=1}^N),
\]
therefore the generating functions are

\[
G^{U(N)}(a,b) =\frac{1}{N!}\left(\frac{1-b}{1-a}\right)^N
\int_0^{2\pi}\prod_i\frac{d\alpha_i}{2\pi} \prod_{i \ne
j}(1-\frac{z_i}{z_j})\frac{(1-b\frac{z_i}{z_j})}{(1-a\frac{z_i}{z_j})},
\]
\[
G^{SU(N)}(a,b) =\frac{1}{N!}\left(\frac{1-b}{1-a}\right)^{N-1}
\int_0^{2\pi}\prod_i\frac{d\alpha_i}{2\pi}\delta(\alpha_N)
\prod_{i \ne
j}(1-\frac{z_i}{z_j})\frac{(1-b\frac{z_i}{z_j})}{(1-a\frac{z_i}{z_j})},
\]
where in the last integral we changed variables  $z_i \rightarrow
z_i/\prod_{j=1}^N z_j$, $z_N\rightarrow \prod_{j=1}^N z_j $.

\section{Appendix B - Explicit calculation of $G^{U(N)}(a,0)$ }

Here we evaluate the integral (\ref{8}) explicitly for $b=0$. It can be
done with use of the Cauchy determinant formula  \footnote{ The
general form of Cauchy determinant formula is
\[
det\left( \frac{1}{z_i-x_j}\right)= \frac{\prod_{i<j}(z_i-
z_j)(x_i-x_j)}{\prod_{i,j}(z_i- x_j)},
\]
which for $x_i=az_i$ yields (\ref{56}).
 }
\begin{equation}
det\left( \frac{1}{z_i-az_j}\right)=
\frac{a^\frac{N(N-1)}{2}}{(a-1)^N}\prod_i \frac{1}{z_i}\prod_{i<j}
\frac{(z_i- z_j)^2}{(z_i-a z_j)(z_j-a z_i)}, \label{56}
\end{equation}
which for $U(N)$ gives
\begin{equation}
G^{U(N)}(a,0) =\frac{1}{N!}\frac{(-1)^N}{a^\frac{N(N-1)}{2}}\int_{\mid z_i \mid=1}
\prod_{k=1}^N \frac{dz_k}{2\pi i} det\left(
\frac{1}{z_i-az_j}\right). \label{57}
\end{equation}
The determinant under the integral (\ref{57})  can be expressed as a sum
over cycles. The integration over each cycle can be done
separately and it gives the factor $\frac{1}{1-a^k}$, i.e.

\[
\int_{\mid z_i \mid=1}  \prod_{i=1}^k \frac{dz_i}{2\pi i}
\frac{1}{z_1-az_2}\frac{1}{z_2-az_3}\ldots
\frac{1}{z_{k-1}-az_k}\frac{1}{z_k-az_1}=\frac{1}{1-a^k},
\]
therefore we obtain

\begin{equation}
\int_{\mid z_i \mid=1}  \prod_{k=1}^N \frac{dz_k}{2\pi i}
det\left( \frac{1}{z_i-az_j}\right)=\sum_{i_1+2i_2+\ldots+Ni_n=N}
(-1)^{\sum_{k=1}^n i_k}L_{i_1 \ldots i_N}\prod_{k=1}^N\frac{1}{(1-a^k)^{i_k}}, \label{58}
\end{equation}

\noindent where $L_{i_1 \ldots i_N}$ is  the number of different
permutations with the same cycle structure given by the partition
$(1^{i_1} \ldots N^{i_N})$ i.e. $L_{i_1 \ldots
i_N}=N!/\prod_{k=1}^N k^{i_k}i_k !$. The right hand side of (\ref{58}) is
in fact very simple due to the Cayley identity \footnote{The Cayley identity by definition is
\[
\sum_{i_1+2i_2+\ldots+Ni_n=N}  L_{i_1 \ldots
i_N}\prod_{k=1}^N\frac{1}{(1-a^k)^{i_k}}= N!\prod_{k=1}^N\frac{1}{(1-a^k)},
\]
which for $a \rightarrow 1/a$ yields (\ref{59}).
}

\begin{equation}
\sum_{i_1+2i_2+\ldots+Ni_n=N} (-1)^{\sum_{k=1}^n i_k} L_{i_1 \ldots
i_N}\prod_{k=1}^N\frac{1}{(1-a^k)^{i_k}}=(-1)^N a^\frac{N(N-1)}{2} N!\prod_{k=1}^N\frac{1}{(1-a^k)}. \label{59}
\end{equation}

\noindent It can be proven most efficiently with use of the Bell polynomials
\cite{bell}. Therefore we finally obtain (\ref{9}).

For $SU(N)$ the only difference is that $Tr(A)=0$ hence in Eqn.(\ref{9}) there
is no $1/(1-a)$ factor, i.e.
\[
G^{SU(N)}(a,0)=(1-a)G^{U(N)}(a,0)=\sum_{n_B=0}^{\infty} a^{n_B}q_N(n_B),
\]
where $q_N(n_B)$ is the number of partitions of $n_B$ into numbers $2,3,\ldots,N$.

The generating function $G^{SU(N)}(a,b)$ clearly
have the form
\[
G^{SU(N)}(a,b)=\left(\prod_{k=2}^N\frac{1}{(1-a^k)}\right)\sum_{i=0}^{N^2-1} (-1)^ib^ic^{SU(N)}_i(a),
\]
where  $c^{SU(N)}_i(a)$ are polynomials in
variable $a$.

\section{Appendix C - Examples of polynomials $c^{U(N)}$ and $c^{SU(N)}$}

Here we list the polynomials $c^{U(N)}$ and $c^{SU(N)}$ for
$N=2,3,4$. They can be obtained from equation (\ref{8}) using some symbolic program, e.g. Mathematica, to evaluate the corresponding residues. For $N=2$ they are
\[
c^{U(2)}_0=1, \ \ \ \  c^{U(2)}_1=1+a, \ \ \ \  c^{U(2)}_2=2a, \ \ \
\ c^{U(2)}_i=c^{U(2)}_{4-i}.
\]
and
\[
c^{SU(2)}_0=1, \ \ \ \  c^{SU(2)}_1=a,\ \ \ \ c^{SU(2)}_i=c^{SU(2)}_{3-i}.
\]
For $N=3$ they are
\[
c^{U(3)}_0=1, \ \ \ \ c^{U(3)}_1=1+a+a^2, \ \ \ \ c^{U(3)}_2=2a+2a^2+2a^3,
\]
\[
c^{U(3)}_3=1+2a+3a^2+5a^3+a^4, \ \ \ \ c^{U(3)}_4=1+3a+6a^2+5a^3+3a^4,\ \ \ \ c^{U(3)}_i=c^{U(3)}_{9-i}.
\]
and
\[
c^{SU(3)}_0=1, \ \ \ \ c^{SU(3)}_1=a+a^2, \ \ \ \ c^{SU(3)}_2=a+a^2+2a^3,
\]
\[
c^{SU(3)}_3=1+a+2a^2+3a^3+a^4, \ \ \ \ c^{SU(3)}_4=2a+4a^2+2a^3+2a^4 \ \ \ \ c^{SU(3)}_i=c^{SU(3)}_{8-i}.
\]
For $N=4$ they are
\[
c^{U(4)}_0=1, \ \ \ \ c^{U(4)}_1=1+a+a^2+a^3, \ \ \ \ c^{U(4)}_2=2a+2a^2+4a^3+2a^3+2a^5,
\]
\[
c^{U(4)}_3=1+2a+4a^2+8a^3+8a^4+7a^5+5a^6+a^7, \ \ \ \ c^{U(4)}_4=1+3a+9a^2+13a^3+19a^4+17a^5+18a^6+7a^7+3a^8,
\]
\[c^{U(4)}_5=1+4a+11a^2+22a^3+33a^4+38a^5+34a^6+23a^7+11a^8+3a^9,
\]
\[
c^{U(4)}_6=1+5a+12a^2+33a^3+45a^4+62a^5+55a^6+45a^7+22a^8+11a^9+a^{10},
\]
\[
c^{U(4)}_7=1+5a+16a^2+37a^3+59a^4+75a^5+77a^6+60a^7+37a^8+17a^9+4a^{10},
\]
\[
c^{U(4)}_8=2+4a+18a^2+36a^3+68a^4+78a^5+86a^6+64a^7+46a^8+18a^9+6a^{10}.
\]
and
\[
c^{SU(4)}_0=1, \ \ \ \ c^{SU(4)}_1=a+a^2+a^3, \ \ \ \ c^{SU(4)}_2=a+a^2+3a^3+2a^4+2a^5,
\]
\[
c^{SU(4)}_3=1+a+3a^2+5a^3+6a^4+5a^5+5a^6+a^7, \ \ \ \ c^{SU(4)}_4=2a+6a^2+8a^3+13a^4+ 12a^5+13a^6+6a^7+3a^8,
\]
\[c^{SU(4)}_5=1+2a+5a^2+14a^3+20a^4+26a^5+21a^6+17a^7+8a^8+3a^9,
\]
\[
c^{SU(4)}_6=3a+7a^2+19a^3+25a^4+36a^5+34a^6+28a^7+14a^8+8a^9+a^{10},
\]
\[
c^{SU(4)}_7=1+2a+9a^2+18a^3+34a^4+39a^5+43a^6+32a^7+23a^8+9a^9+3a^{10}.
\]

\section{Appendix D - Linear dependence of the trace operators for the $SU(3)$ group }

Here we derive the linear dependence of 17 trace operators listed in
subsection 2.1.1. Our starting point is the equation

\begin{equation}
a^{\dagger}a^{\dagger}f^{\dagger}+
f^{\dagger}a^{\dagger}a^{\dagger}+a^{\dagger}f^{\dagger}a^{\dagger}=
\frac{1}{2}(a^{\dagger}a^{\dagger})f^{\dagger}+(a^{\dagger}f^{\dagger})a^{\dagger}+(a^{\dagger}a^{\dagger}f^{\dagger}). \label{62}
\end{equation}
It is the source of the following
relations.

\vspace{0.5cm} \noindent Multiplying (\ref{62}) from the right hand side
by $f^{\dagger}f^{\dagger}$ and taking the trace gives
\[
(a^{\dagger}f^{\dagger}a^{\dagger}f^{\dagger}f^{\dagger})=
\frac{1}{2}(a^{\dagger}a^{\dagger})(f^{\dagger}f^{\dagger}f^{\dagger})+
(a^{\dagger}f^{\dagger})(a^{\dagger}f^{\dagger}f^{\dagger})-
2(a^{\dagger}a^{\dagger}f^{\dagger}f^{\dagger}f^{\dagger}).
\]
Therefore, we may neglect, e.g. $(a^{\dagger}f^{\dagger}a^{\dagger}f^{\dagger}f^{\dagger})$.

\vspace{0.5cm} \noindent Multiplying (\ref{62}) from the right hand side
by $f^{\dagger}f^{\dagger}$ and from the left hand side by
$a^{\dagger}$ and then taking the trace gives
\[
(a^{\dagger}a^{\dagger}f^{\dagger}a^{\dagger}f^{\dagger}f^{\dagger})+
(a^{\dagger}f^{\dagger}a^{\dagger}a^{\dagger}f^{\dagger}f^{\dagger})=
(a^{\dagger}f^{\dagger})(a^{\dagger}a^{\dagger}f^{\dagger}f^{\dagger})+
(a^{\dagger}a^{\dagger}f^{\dagger})(a^{\dagger}f^{\dagger}f^{\dagger})-
\frac{1}{3}(a^{\dagger}a^{\dagger}a^{\dagger})(f^{\dagger}f^{\dagger}f^{\dagger}),
\]
where the Cayley-Hamilton theorem for $a^{\dagger}$ matrices was also used.

\vspace{0.5cm} \noindent Multiplying (\ref{62}) from the right hand side
by $f^{\dagger}a^{\dagger}f^{\dagger}$ and taking the trace gives
\[
(a^{\dagger}a^{\dagger}f^{\dagger}f^{\dagger}a^{\dagger}f^{\dagger})+
(a^{\dagger}f^{\dagger}a^{\dagger}f^{\dagger}a^{\dagger}f^{\dagger})+
(a^{\dagger}a^{\dagger}f^{\dagger}a^{\dagger}f^{\dagger}f^{\dagger})=
\frac{1}{2}(a^{\dagger}a^{\dagger})(a^{\dagger}f^{\dagger}f^{\dagger}f^{\dagger})-
(a^{\dagger}a^{\dagger}f^{\dagger})(a^{\dagger}f^{\dagger}f^{\dagger}).
\]
From the two above equations it follows that
$(a^{\dagger}f^{\dagger}a^{\dagger}f^{\dagger}a^{\dagger}f^{\dagger})$
can be expressed in terms of multiple trace operators. Also, we may
neglect, e.g.
$(a^{\dagger}a^{\dagger}f^{\dagger}a^{\dagger}f^{\dagger}f^{\dagger})$.

\vspace{0.5cm} \noindent Multiplying (\ref{62}) from the right hand side
by $ a^{\dagger}a^{\dagger}f^{\dagger}f^{\dagger}$ and taking the
trace gives
\[
(a^{\dagger}a^{\dagger}f^{\dagger}a^{\dagger}a^{\dagger}f^{\dagger}f^{\dagger})=
\frac{1}{2}(a^{\dagger}f^{\dagger})(a^{\dagger}a^{\dagger})(a^{\dagger}f^{\dagger}f^{\dagger})+
(a^{\dagger}a^{\dagger}f^{\dagger})(a^{\dagger}a^{\dagger}f^{\dagger}f^{\dagger})
-\frac{2}{3}(a^{\dagger}a^{\dagger}a^{\dagger})(a^{\dagger}f^{\dagger}f^{\dagger}f^{\dagger})
\]
\[
-\frac{1}{2}(a^{\dagger}a^{\dagger})(a^{\dagger}f^{\dagger}a^{\dagger}f^{\dagger}f^{\dagger}),
\]
where the Cayley-Hamilton theorem for $a^{\dagger}$ matrices was
also used. Therefore we neglect
$(a^{\dagger}a^{\dagger}f^{\dagger}a^{\dagger}a^{\dagger}f^{\dagger}f^{\dagger})$.

\vspace{0.5cm} \noindent Multiplying (\ref{62}) from the right hand side
by $f^{\dagger}a^{\dagger}a^{\dagger}f^{\dagger}$ and taking the
trace gives
\[
2(a^{\dagger}a^{\dagger}f^{\dagger}a^{\dagger}a^{\dagger}f^{\dagger}f^{\dagger})+
(a^{\dagger}a^{\dagger}f^{\dagger}a^{\dagger}f^{\dagger}a^{\dagger}f^{\dagger})=
\frac{1}{2}(a^{\dagger}a^{\dagger})(a^{\dagger}a^{\dagger}f^{\dagger}f^{\dagger}f^{\dagger})+
\frac{1}{2}(a^{\dagger}f^{\dagger})(a^{\dagger}f^{\dagger}a^{\dagger}a^{\dagger}f^{\dagger})
\]
\[
-(a^{\dagger}a^{\dagger}f^{\dagger})(a^{\dagger}a^{\dagger}f^{\dagger}f^{\dagger}),
\]
where the Cayley-Hamilton theorem for $a^{\dagger}$ matrices was
again used. From the two above equations it follows that we can also
neglect
$(a^{\dagger}a^{\dagger}f^{\dagger}a^{\dagger}f^{\dagger}a^{\dagger}f^{\dagger})$.

\vspace{0.5cm} \noindent Multiplying (\ref{62}) from the right hand side
by $a^{\dagger}f^{\dagger}a^{\dagger}f^{\dagger}$ and taking the
trace gives
\[
2(a^{\dagger}a^{\dagger}f^{\dagger}a^{\dagger}f^{\dagger}a^{\dagger}f^{\dagger})+
\frac{1}{2}(a^{\dagger}a^{\dagger})(f^{\dagger}a^{\dagger}f^{\dagger}a^{\dagger}f^{\dagger})+
\frac{1}{3}(a^{\dagger}a^{\dagger}a^{\dagger})(f^{\dagger}f^{\dagger}a^{\dagger}f^{\dagger})=
\frac{1}{2}(a^{\dagger}a^{\dagger})(f^{\dagger}a^{\dagger}f^{\dagger}a^{\dagger}f^{\dagger})
\]
\[
+(a^{\dagger}f^{\dagger})(a^{\dagger}a^{\dagger}f^{\dagger}a^{\dagger}f^{\dagger}),
\]
where the Cayley-Hamilton theorem for $a^{\dagger}$ matrices was
used. The operator
$(a^{\dagger}a^{\dagger}f^{\dagger}a^{\dagger}f^{\dagger}a^{\dagger}f^{\dagger})$
was already excluded before hence, from the above identity, it
follows that there is a relation between the multiple trace
operators. Therefore we can neglect one, e.g.
$(a^{\dagger}f^{\dagger})(a^{\dagger}a^{\dagger}f^{\dagger}a^{\dagger}f^{\dagger})$.

\vspace{0.5cm} \noindent Multiplying (\ref{62}) from the right hand side
by $a^{\dagger}a^{\dagger}f^{\dagger}a^{\dagger}f^{\dagger}$ and
taking the trace gives
\[
(a^{\dagger}a^{\dagger}f^{\dagger}a^{\dagger}a^{\dagger}f^{\dagger}a^{\dagger}f^{\dagger})=
-\frac{1}{3}(a^{\dagger}f^{\dagger})(a^{\dagger}a^{\dagger}a^{\dagger})(a^{\dagger}f^{\dagger}f^{\dagger})+
(a^{\dagger}a^{\dagger}f^{\dagger})(a^{\dagger}a^{\dagger}f^{\dagger}a^{\dagger}f^{\dagger})-
\frac{2}{3}(a^{\dagger}a^{\dagger}a^{\dagger})(a^{\dagger}f^{\dagger}a^{\dagger}f^{\dagger}f^{\dagger})
\]
\[
-
\frac{1}{2}(a^{\dagger}a^{\dagger})(a^{\dagger}f^{\dagger}a^{\dagger}f^{\dagger}a^{\dagger}f^{\dagger}),
\]
where the Cayley-Hamilton theorem for $a^{\dagger}$ matrices was
used. Therefore, we neglect the operator
$(a^{\dagger}a^{\dagger}f^{\dagger}a^{\dagger}a^{\dagger}f^{\dagger}a^{\dagger}f^{\dagger})$.

\vspace{0.5cm} \noindent Multiplying (\ref{62}) from the right hand side
by $a^{\dagger}f^{\dagger}a^{\dagger}a^{\dagger}f^{\dagger}$ and
taking the trace gives
\[
2(a^{\dagger}a^{\dagger}f^{\dagger}a^{\dagger}a^{\dagger}f^{\dagger}a^{\dagger}f^{\dagger})+
\frac{1}{2}(a^{\dagger}a^{\dagger})(f^{\dagger}a^{\dagger}f^{\dagger}a^{\dagger}a^{\dagger}f^{\dagger})+
\frac{1}{3}(a^{\dagger}a^{\dagger}a^{\dagger})(f^{\dagger}f^{\dagger}a^{\dagger}a^{\dagger}f^{\dagger})=
\frac{1}{2}(a^{\dagger}a^{\dagger})(f^{\dagger}a^{\dagger}f^{\dagger}a^{\dagger}a^{\dagger}f^{\dagger})
\]
\[
+(a^{\dagger}a^{\dagger}f^{\dagger})(a^{\dagger}f^{\dagger}a^{\dagger}a^{\dagger}f^{\dagger}),
\]
where the Cayley-Hamilton theorem for $a^{\dagger}$ matrices was
used. The
$(a^{\dagger}a^{\dagger}f^{\dagger}a^{\dagger}a^{\dagger}f^{\dagger}a^{\dagger}f^{\dagger})$
is already excluded therefore we can neglect one multitrace
operator, e.g.
$(a^{\dagger}a^{\dagger}f^{\dagger})(a^{\dagger}a^{\dagger}f^{\dagger}a^{\dagger}f^{\dagger})$.

\vspace{0.5cm} \noindent Multiplying (\ref{62}) from the right hand side
by
$a^{\dagger}a^{\dagger}f^{\dagger}a^{\dagger}a^{\dagger}f^{\dagger}$
and taking the trace gives
\[
(a^{\dagger}a^{\dagger}f^{\dagger}a^{\dagger}a^{\dagger}f^{\dagger}a^{\dagger}a^{\dagger}f^{\dagger})+
\frac{1}{2}(a^{\dagger}a^{\dagger})(a^{\dagger}f^{\dagger}a^{\dagger}f^{\dagger}a^{\dagger}a^{\dagger}f^{\dagger})+
\frac{1}{3}(a^{\dagger}a^{\dagger}a^{\dagger})(a^{\dagger}f^{\dagger}f^{\dagger}a^{\dagger}a^{\dagger}f^{\dagger})
\]
\[
+\frac{1}{2}(a^{\dagger}a^{\dagger})(f^{\dagger}a^{\dagger}a^{\dagger}f^{\dagger}a^{\dagger}a^{\dagger}f^{\dagger})
+\frac{1}{3}(a^{\dagger}a^{\dagger}a^{\dagger})(f^{\dagger}a^{\dagger}f^{\dagger}a^{\dagger}a^{\dagger}f^{\dagger})=
\frac{1}{2}(a^{\dagger}a^{\dagger})(f^{\dagger}a^{\dagger}a^{\dagger}f^{\dagger}a^{\dagger}a^{\dagger}f^{\dagger})
\]
\[
+\frac{1}{2}(a^{\dagger}f^{\dagger})(a^{\dagger}a^{\dagger})(a^{\dagger}f^{\dagger}a^{\dagger}a^{\dagger}f^{\dagger})+
\frac{1}{3}(a^{\dagger}f^{\dagger})(a^{\dagger}a^{\dagger}a^{\dagger})(f^{\dagger}a^{\dagger}a^{\dagger}f^{\dagger}),
\]
where the Cayley-Hamilton theorem for $a^{\dagger}$ matrices was
also used. It follows that the operator
$(a^{\dagger}a^{\dagger}f^{\dagger}a^{\dagger}a^{\dagger}f^{\dagger}a^{\dagger}a^{\dagger}f^{\dagger})$
can be excluded as well.

There are no other independent  relations following from (\ref{62}) except
from the ones where the Cayley-Hamilton theorem for $a^{\dagger}$
matrices is used. We excluded 9 operators as being  linearly
dependent. The remaining 8 trace operators are indicated in subsection
2.1.1  .

\section{Appendix E - Examples of generating functions $G^{SU(N), d}(a,b)$}

Below we list the generating functions $G^{SU(N), d}(a,b)$, for
$d=3,5,9$, $N=2,3$. The case of $U(N)$ group is obtained from the
identity
\[
G^{U(N), d}(a,b)=\frac{(1-b)^{d-1}}{(1-a)^d}G^{SU(N), d}(a,b).
\]

\noindent For $d=3$, $N=2$ we have
\[
G^{SU(2), d=3}(a,b)=\frac{1}{(1-a)(1-a^2)^5}\sum_{i=0}^6b^ic^{SU(2), d=3}_i,
\]
where $c^{SU(2), d=3}_i=c^{SU(2), d=3}_{6-i}$ and
\[
c^{SU(2), d=3}_0=1-a+a^2, \ \ \ \ c^{SU(2), d=3}_1=6a, \ \ \ \ c^{SU(2), d=3}_2=1+8a+7a^2+2a^3-2a^4-a^5,
\]
\[
c^{SU(2), d=3}_3=4+2a+16a^2+4a^3-4a^4-2a^5.
\]

\noindent For $d=5$, $N=2$ we have
\[
G^{SU(2), d=5}(a,b)=\frac{1}{(1-a)^3(1-a^2)^9}\sum_{i=0}^{12}(-1)^i c^{SU(2), d=5}_i b^i,
\]
\[
c^{SU(2), d=5}_0=1-3a+9a^2-9a^3+9a^4-3a^5+a^6.
\]
We will not list the rest of $c_i$'s since there are many of them and they become more complicated.

\noindent For $d=9$, $N=2$ we have
\[
G^{SU(2), d=9}(a,b)=\frac{1}{(1-a)^7(1-a^2)^{17}}\sum_{i=0}^{24}(-1)^i c^{SU(2), d=9}_i b^i,
\]
\[
c^{SU(2), d=9}_0=1-7a+49a^2-147a^3+441a^4-735a^5+1225a^6-1225a^7+
+1225a^8-735a^9
\]
\[
+441a^{10}-147a^{11}+49a^{12}-7a^{13}+a^{14}.
\]

\noindent For $d=3$, $N=3$ we have
\[
G^{SU(3), d=3}(a,b)=\frac{1}{(1-a)(1-a^2)^8 (1-a^3)^7}\sum_{i=0}^{16}(-1)^i c^{SU(3), d=3}_i b^i,
\]
\[
c^{SU(3), d=3}_0=1-a-2a^2+6a^3+6a^4-9a^5+a^6+17a^7+a^8-9a^9+6a^{10}+6a^{11}-2a^{12}-a^{13}+a^{14}.
\]

\noindent For $d=5$, $N=3$ we have
\[
G^{SU(3), d=5}(a,b)=\frac{1}{(1-a)^3(1-a^2)^{16} (1-a^3)^{13}}\sum_{i=0}^{32}(-1)^i c^{SU(3), d=5}_i b^i,
\]
\[
c^{SU(3), d=5}_0=1-3a+2a^2+34a^3-4a^4-18a^5+421a^6+ 624a^7+251a^8+2107a^9+5377a^{10}+4766a^{11}
\]
\[
+6384a^{12}+16031a^{13}+19327a^{14}+14592a^{15}+21381a^{16}+29839a^{17}+21381a^{18}+14592a^{19}+19327a^{20}
\]
\[
+16031a^{21}+6384a^{22}+4766a^{23}
+5377a^{24}+2107a^{25}+251a^{26}+624a^{27}+421a^{28}-18a^{29}-4a^{30}+34a^{31}
\]
\[
+2a^{32}-3a^{33}+a^{34}.
\]

\noindent For $d=9$, $N=3$ we have
\[
G^{SU(3), d=9}(a,b)=\frac{1}{(1-a)^6(1-a^2)^{32} (1-a^3)^{25}}\sum_{i=0}^{64}(-1)^i c^{SU(3),d=9}_i b^i,
\]
where $c^{SU(3),d=9}_0$ is of order 74.

\section{Appendix F - Number of gauge singlets with given angular momentum   }

Here we discuss the character method applied to sectors with fixed
angular momentum. The projection to sectors with fixed angular
momentum $j$ is due to the decomposition
\[
V=Sym(\otimes_{k=1}^{n_B}A_k^{j=1})\times Alt(\otimes_{l=1}^{n_F}F_l^{j=1/2}),
\]
where $A_k^{j=1}$, $F_l^{j=1/2}$ are  vector spaces spanned by
$a^{j=1}_k\mid 0 \rangle$, and $f^{j=1/2}_l\mid 0 \rangle$ where
operators $a^{j=1}_k$, $f^{j=1/2}_l$ are assumed to carry $SO(3)$ spin $1$
and $1/2$ respectively. Therefore the dimensions of subspaces with
angular momentum $j$ are
\[
D^{U(N), \ d, \ j}_{n_B  \ n_F} = \int d\mu_{SO(d)}
\chi^{SO(d),j}   \int d\mu_{U(N)}
\chi^{\{n_B\}}_{Sym}(R^{j=1}_B)
\chi^{[n_F]}_{Alt}(R^{j=1/2}_F),
\]
where $d\mu_{SU(N)}$ and  $d\mu_{SO(d)}$  are $SU(N)$ and $SO(d)$
invariant measures. We will restrict to the $d=3$ case hence we take

\[
d\mu_{SO(3)}=\frac{1}{\pi}\sin^2 \frac{\beta}{2}d\beta,  \ \ \ \
\beta \in [0,2\pi], \ \ \ \ \int d\mu_{SO(3)} =1.
\]
$R^{j=1}_B$, $R^{j=1/2}_B$ are the  adjoint representation of
$SU(N)$ and $j=1$, $j=1/2$ representations of $SO(d)$ respectively,
i.e.
\[
\chi(R^{j=1}_{B})=\chi(R^{SO(d), \ j=1}) \chi(R^{SU(N), \ j=1} ),
\]
\[
\chi(R^{j=1/2}_{F})=\chi(R^{SO(d), \ j=1/2}) \chi(R^{SU(N), \ j=1}
).
\]
For $d=3$ we have
\[
\chi^{SO(3),j}(\alpha)=\frac{\sin(j+\frac{1}{2})\alpha}{\sin\frac{1}{2}\alpha}=\sum_{k=-j}^{k=j}t^k, \ \ \ \ t=e^{i\alpha},
\]
therefore for $SU(2)$ gauge group we obtain
\[
\chi(R^{j=1}_{B})=\chi(R^{SO(3), \ j=1}) \chi(R^{SU(2), \ j=1}
)=(1+2\cos\beta)(1+2\cos\alpha),
\]
\[
\chi(R^{j=1/2}_{F})=\chi(R^{SO(3), \ j=1/2}) \chi(R^{SU(2), \ j=1}
)=2\cos\frac{\beta}{2}(1+2\cos\alpha).
\]

The explicit calculation of the generating  function for $D^{SU(2), \
d=3, \ j}_{n_B  \ n_F}$ is now straightforward. Below we perform the
calculation for the purely bosonic sector, i.e. we
evaluate
\[
G(a,c)=\sum_{n_B=0}^{\infty}\sum_{j=0}^{\infty}D^{SU(2), \ d=3, \ j}_{n_B  \ 0}a^{n_B}c^j.
\]
Using the same conventions and techniques as in Appendix A we
perform the sum over $n_B$. We have
\[
G(a,c)=\sum_{j=0}^{\infty}c^j\int d \mu_{SO(3)}d \mu_{SU(2)}  \left(
\sum_{k=-j}^{k=j}t^k \right) F(a,z;c,t), \ \ \ \ z=e^{i\alpha}, \ \ \ \ t=e^{i\beta},
\]
where
\[
F(a,z;c,t)=\frac{t^3z^3}{(1-a)(1-at)(1-az)(1-atz)(a-t)(a-z)(a-tz)(at-z)(t-az)}.
\]
Now the sum over $j$ is also possible
and the evaluation of the resulting integral gives
\[
G(a,c)=\frac{1-a^2c+a^4c^2}{(1-a^2)(1-a^3)(1-a^4)(1-a^2c)(1-a^2c^2)}.
\]

\end{document}